# Perpendicular separations of a binary mixture under van der Waals confinement

Kui Lin*
Department of Civil and Environmental Engineering, The Hong Kong Polytechnic University, Hong Kong, China.

We investigated the dynamics of a binary mixture confined within van der Waals (vdW) walls using molecular dynamics simulations. We discovered a novel phenomenon named perpendicular separations of two phases (PSTP). In the initial stage, central water molecules diffused, subsequently condensing symmetrically within the confinement's mid-plane. In the later stage, as water droplets nucleate and grow, the resin separates perpendicularly into two films due to the action of bubblers and vdW walls, resulting in a hollow nanochannel. The mechanisms and conditions underlying PSTS are discussed. The results indicate that the concentration ($C$) of resin in the middle region is linearly decreased with temporal power ($C(t,T) \propto a(T)t^{1/3}$). We propose a new mechanism for stabilizing nanochannels and films: dynamic "soft pillars" that prevent Rayleigh-like instability. Our findings could shed light on the manufacture of nanofilms and organic nanochannels, which could help advance bio-detection and energy fields.

*Introduction.* The phase separation of polymeric mixtures has been a focus of intensive research over the past four decades, primarily due to its technological significance in tailoring the properties of new polymeric materials [1]. The wetting dynamics and phase separation of a binary mixture confined in one-/two-dimensional (1-D/2-D) capillaries have been extensively studied using experimental and theoretical methods [2-5]. These studies have highlighted the pivotal role of hydrodynamics in the evolution of phase separation patterns, such as bicontinuous and droplet patterns. Owing to the structural symmetry, liquids distribute symmetrically within the confined space. This indicates that the more wettable liquid tends to adhere to walls and the less wettable liquid is likely to distribute in the central region. Rayleigh-like instability [4] can cause bridge formation between two wetting layers in a 1D capillary. However, a more stable wetting layer can be formed in a 2D capillary since small perturbations consistently increase the area of the planar interface [3]. By analyzing the free energies of the configurations, two possible final configurations, namely partial wetting and complete wetting, are formed, and the morphological transition occurs at $\phi_{\text{less}} = \sigma/\Delta\gamma$. Here $\phi_{\text{less}}$ represents the volume fraction of the less wettable phase, $\sigma$ denotes the interfacial tension between the two phases, and $\Delta\gamma$ signifies the difference in the interfacial tensions between the two phases with the wall [3].

The phenomena described above all occur at the mesoscale, where classical capillary action is the primary driver of phase separation and wetting dynamics. However, generally, the driving pressure, $p$, of fluid in nanochannels is a combination of the classical capillary pressure, $p_{\text{cap}}$, and the disjoining pressure, $\Pi$ [6-8], represented as $p = p_{\text{cap}} + \Pi$. The disjoining pressure comprises van der Waals (vdW) contributions, $\Pi_{\text{vdW}} = \mathrm{A}/6\pi h^3$, and entropic terms. The former predominates, with A representing the Hamaker constant, and $h$ representing the film thickness of the liquid, which drives the behavior of liquids at the nanoscale. Experimental evidence has confirmed that vdW forces play a pivotal role, triggering the formation of a thin polymer resin film on the substrate [9]. Conversely, the vdW force serves as another stabilizing factor [5], capable of reducing Rayleigh-like instability. Consequently, the creation of nano-confinement structures dominated by vdW interactions could potentially allow for the formation of stable nano-films through wall attraction.

Regarding the influence of the attraction wall, both experimental and theoretical studies have elucidated the laws governing phase separation in the near-surface region [10-15]. For instance, the different attraction of two components has been observed to cause anisotropic evolution of phase separation and oscillatory composition profiles [11, 15]. Meanwhile, the temporal dependence of spinodal decomposition near a wall, including the growth of the wetting layer thickness that follows with a temporal power $t^{1/3}$, has also been explored in various theoretical studies [12-15]. These studies provide ample experimental evidence and insight into the mechanisms behind thin film growth on wall surfaces.

Nonetheless, the majority of research on binary mixture phase separation in 1-/2-D capillaries and film formation on surfaces, including the aforementioned examples, focuses on micron or submicron scales. To the best of the author's knowledge, the phase separation of a binary mixture under nanoscale vdW confinement remains largely unexplored. This is likely due to the limitation of the experimental techniques, such as the nano-*in situ* methods, and the challenge of fabricating controllable vdW materials. However, these barriers are gradually being overcome with technological advancements. Over the past two decades, 2D materials and vdW heterostructures have been extensively studied [16-18], showcasing their significant potential for practical applications [18, 19]. As a result, advanced nano-experimental techniques and nanostructure materials have opened up research possibilities for nanoconfined liquids [8, 20, 21]. Therefore, as depicted in Fig. 1(a), uniting a binary mixture with vdW confinement is technically feasible, yet the phase separation process in vdW confinement has not been intuitively investigated.

* Corresponding author
Email addresses: kui-cee.lin@polyu.edu.hk, kui.lin.phys@gmail.com

This paper presents the first investigation of phase separation in binary liquids (i.e., water and epoxy resin) under vdW confinement using molecular dynamics (MD) simulations. We observed bulk water diffusion followed by coarsening, a process analogous to the experimental phenomena known as Ostwald ripening. Concurrently, the resin separated to form thin films on the vdW walls, measuring 0.96 nm and 0.84 nm in thickness at 345 K and 375 K, respectively. We discovered a novel phenomenon we have termed 'perpendicular separations of two phases' (PSTP). Ultimately, a nanochannel with a width of approximately between 1 and 2 nm was formed. The evolution profiles of the two phases are detailed, followed by discussions of their associated kinetical laws. The results indicate that the epoxy's decaying velocity in the central region follows a $t^{1/3}$ temporal power-law during the initial ~86% of the decay process, after which the decay velocity transitions to a slow linear relationship with time ($\propto t$). This study presents evidence and a new mechanism for the formation of stable nano-films and organic nano-channels, which have potential applications in bio-detection, medical testing, and energy fields.

*Methodology*. Figure 1(a) illustrates the schematic of the initial state of the binary mixture, i.e., water/epoxy resin, within the vdW confinement. Figure 1(b), representing a slice, display each component of the system. The size of the model is 20×20×4 nm. Initially, bulk water molecules are centrally located within the confined space, forming a cylinder surrounded by epoxy molecules (diglycidyl ether of bisphenol A). The densities of the two liquids mimic their real-world conditions at normal temperature and pressure. The CVFF forcefield [22] was used for the epoxy, and the TIP3P model was used for water molecules. Nonbonding interactions were calculated using the standard 12/6 Lennard-Jones (LJ) potential with a cutoff of 12 Å and Coulombic pairwise interactions with a cutoff of 10 Å. The LJ93 potential, i.e., $U(R) = \epsilon[2/15(\sigma/R)^9 - (\sigma/R)^3]$, with $\epsilon = 0.5$ kcal/mol, $\sigma = 3$ Å, and a cutoff of 10 Å, was used to calculate the vdW interaction energies between the atoms and the vdW walls. Interatomic interactions were calculated according to the Lorentz–Berthelot (LB) rule. Therefore, the Hamiltonian for the confined system includes the kinetic energy $E_{\mathrm{ki}}$, intermolecular interactions $U_{\mathrm{ij}}(\boldsymbol{r}_{\mathrm{ij}})$, liquid-wall interaction $U_{\mathrm{ik}}(\boldsymbol{R}_{\mathrm{ik}})$, i.e., $H = E_{\mathrm{ki}} + \sum_{\mathrm{i,j}} U_{\mathrm{ij}}(\boldsymbol{r}_{\mathrm{ij}}) + \sum_{\mathrm{i,k}} U_{\mathrm{ik}}(\boldsymbol{R}_{\mathrm{ik}})$. All simulations were conducted using LAMMPS [23]. We chose the NPT ensemble with Nose/Hoover method to regulate the target temperatures. The lateral pressure was maintained at 1 atm, and the timestep was set to 1 fs.

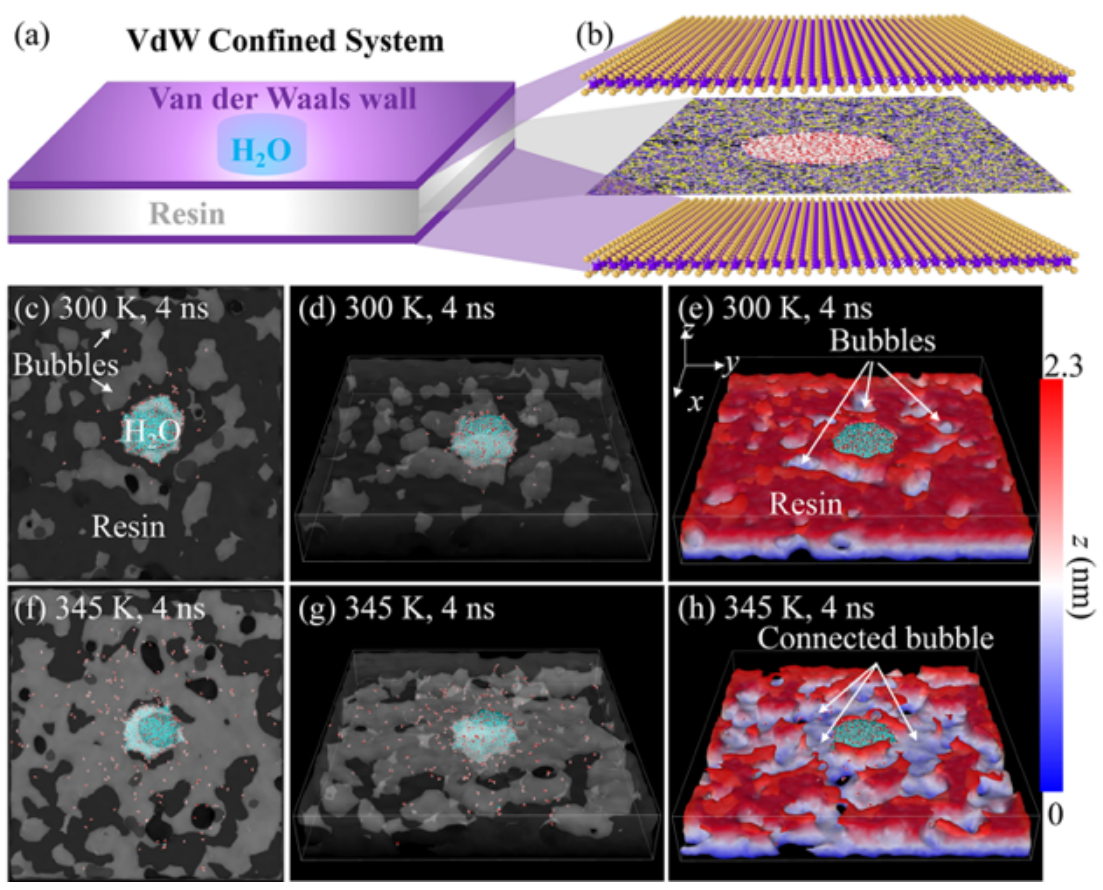

FIG. 1. (a) A schematic of the vdW confined system; (b) An example of the components of the vdW confined system: two-dimensional materials (e.g., $MoS_2$), a binary mixture (e.g., water/resin); Top view (c), oblique view (d), and middle section of water/resin in vdW confinement after 4 ns at 300 K (e), as well as (f-h) for the case at 345 K.

Temperature and mixture concentration are crucial determinants in the phase separation process. Our initial simulations involved the diffusion processes of a water cylinder with a 4 nm diameter at 300 K and 345 K. The internal configuration of the binary mixture at 300 K after 4 ns is shown in Fig. 1(c,d), where dark and grey correspond to resin and bubble/cavity, respectively. Under these conditions, many bubbles remain unconnected as shown in Fig. 1(e). Upon elevation of the temperature to 345 K, a higher number of water molecules infiltrated the resin, leading to an increase in bubble formation, as shown in Fig. 1(f, g). However, the water concentration was not sufficiently high to enable coarsening of the diffused water and the formation of a fully interconnected nano-channel (refer to Fig. 1(h)). Consequently, we enlarged the diameter of the water cylinder to 8 nm to enhance the concentration of water, while maintaining the temperature at 345 K. Quasi-symmetric diffusion and coarsening ( i.e., Ostwald ripening [24]) were observed as shown in Fig. 2(a). Figure 2(b) shows the radial distribution of water molecules originating from the center of the cylinder water. This demonstrates the quasi-symmetric nature and identifies the onset of the coarsening position to be 7 to 7.5 nm distant from the central water source.

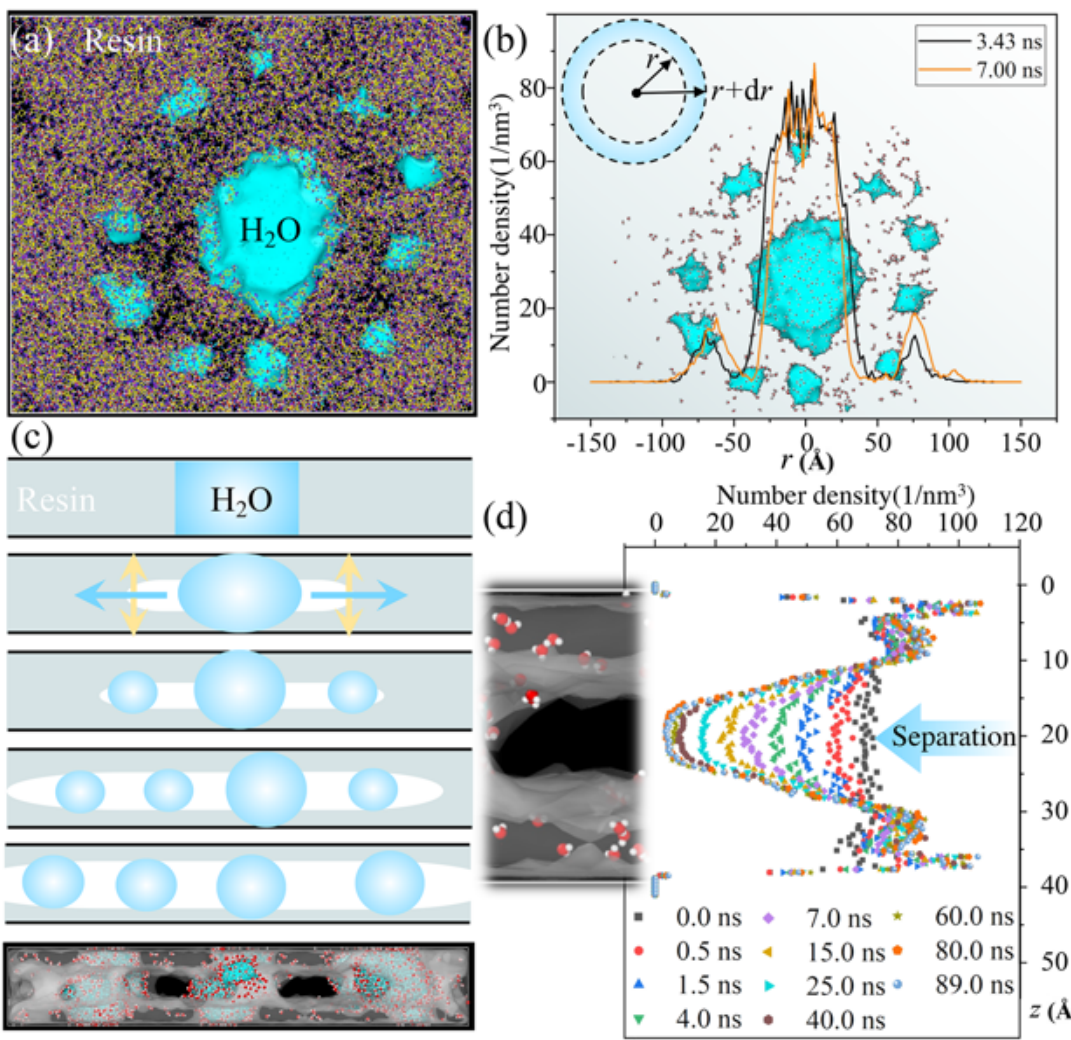

FIG. 2. (a) Quasi-symmetric coarsening of water in the binary mixture; (b) Radial distribution of water molecules; (c) Schematic of the perpendicular separations of two phases; (d) Evolution of the resin distribution in the direction perpendicular to the wall during separation.

Besides the planar diffusion and coarsening of water, the resin also separates in the thickness dimension. This phenomenon, which we term the Perpendicular Separations of Two Phases (PSTP), is shown in Fig. 2 (c). Given that the intermediate water acts as a source, enabling continuous injection of water, the PTST process facilitates the continuous expansion of the film or channel to a desired size, potentially offering significant value in engineering applications. Figure 2(d) illustrates the evolution of resin concentration across the thickness dimension. Figure 3 shows the entirety of the PSTP at 345 K, visibly demonstrating Ostwald ripening and resin separation. The dark region signifies the resin, whereas the gray region represents the bubble resulting from resin separation during water molecules diffusion. As shown in Fig. 3(a-h), the initial stage of coarsening exhibits quasi-symmetrical configurations owning to the structure symmetry. Subsequently, the size of coarsened droplets tends toward homogeneity (with a diameter spanning 2 to 3 nm), and they randomly meander within the confinement, as shown in Fig. 3(i). Consequently, dynamically stable separated phases are eventually formed.

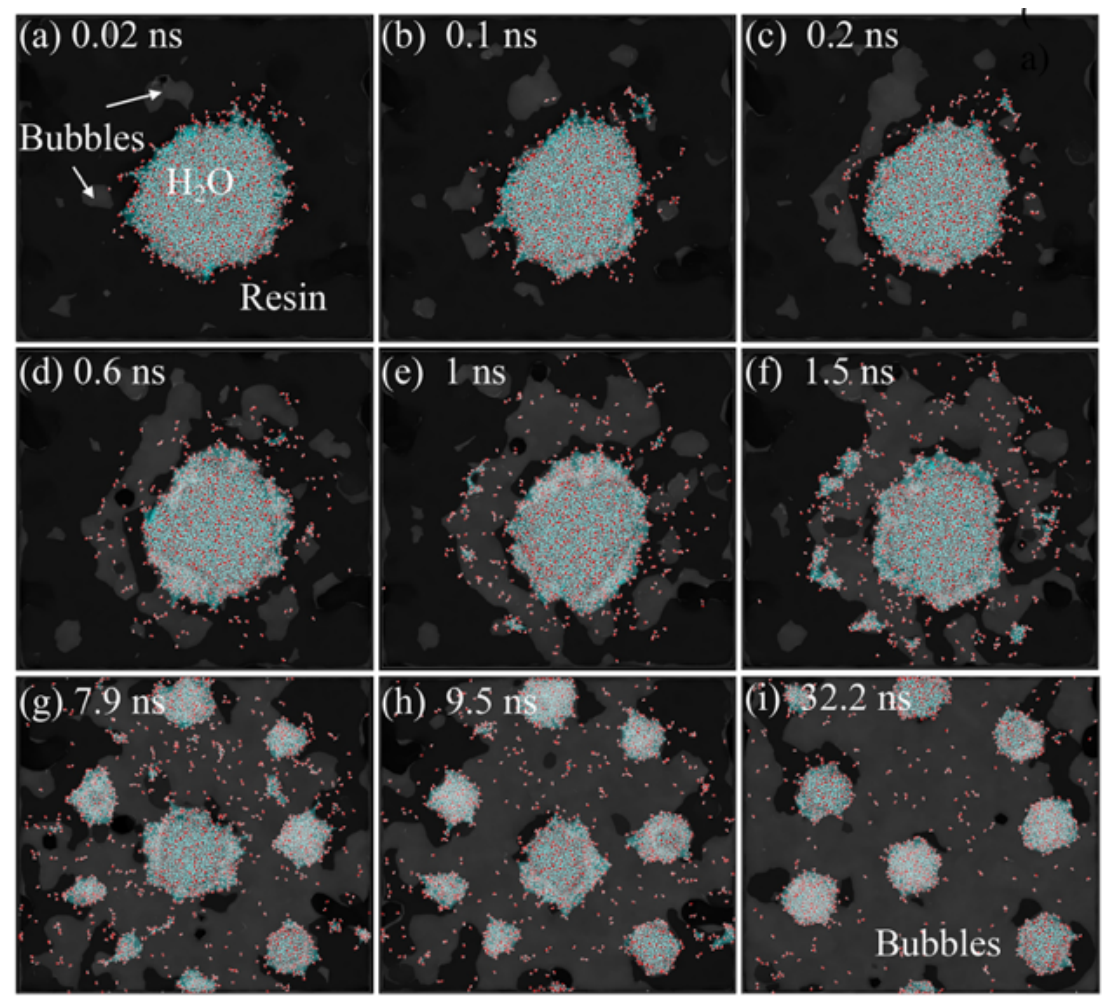

FIG. 3. Ostwald ripening: Details of bulk water diffusing, nucleating, and growing of binary mixture under vdW confinement at a 345 K.

To evaluate the influence of temperature on the PSTP, we simulated the equilibrium process of the identical system at 375 K and subsequently analyzed the morphology of the internal structure. The system reaches equilibrium in approximately 80 ns at 345 K and 40 ns at 375 K, respectively. Consequently, a temperature increase can hasten the PSTP process. Figure 4(a-c, d-f) presents the equilibrium patterns at two distinct temperatures, indicating that a temperature rise could augment the channel's connection rate and reduce the thickness of the resin film on vdW walls (see Fig. 4(c, f)). As a result of thermal disturbance and reduced thickness, additional voids are formed in the film, as shown in Fig. 4(f). This implies that in order to minimize the formation of voids within the film, the operational temperature must be maintained below a certain threshold. To identify the film thickness of the organic nanochannels (see Fig. 4(g, i)), we calculated the number density of atoms of epoxy resin in the thickness direction, thereby obtaining distribution profiles of resin. The thickness ($d$) of the resin films can be determined based on the crossover of resin profiles at initial and equilibrium states, as illustrated in Fig. 4(h, j). Thus, the thicknesses are 0.96 nm and 0.84 nm at 345 K and 375 K, respectively. This is due to the combined effects of accelerated thermal motion within the polymer and the weakened van der Waals interactions with the wall.

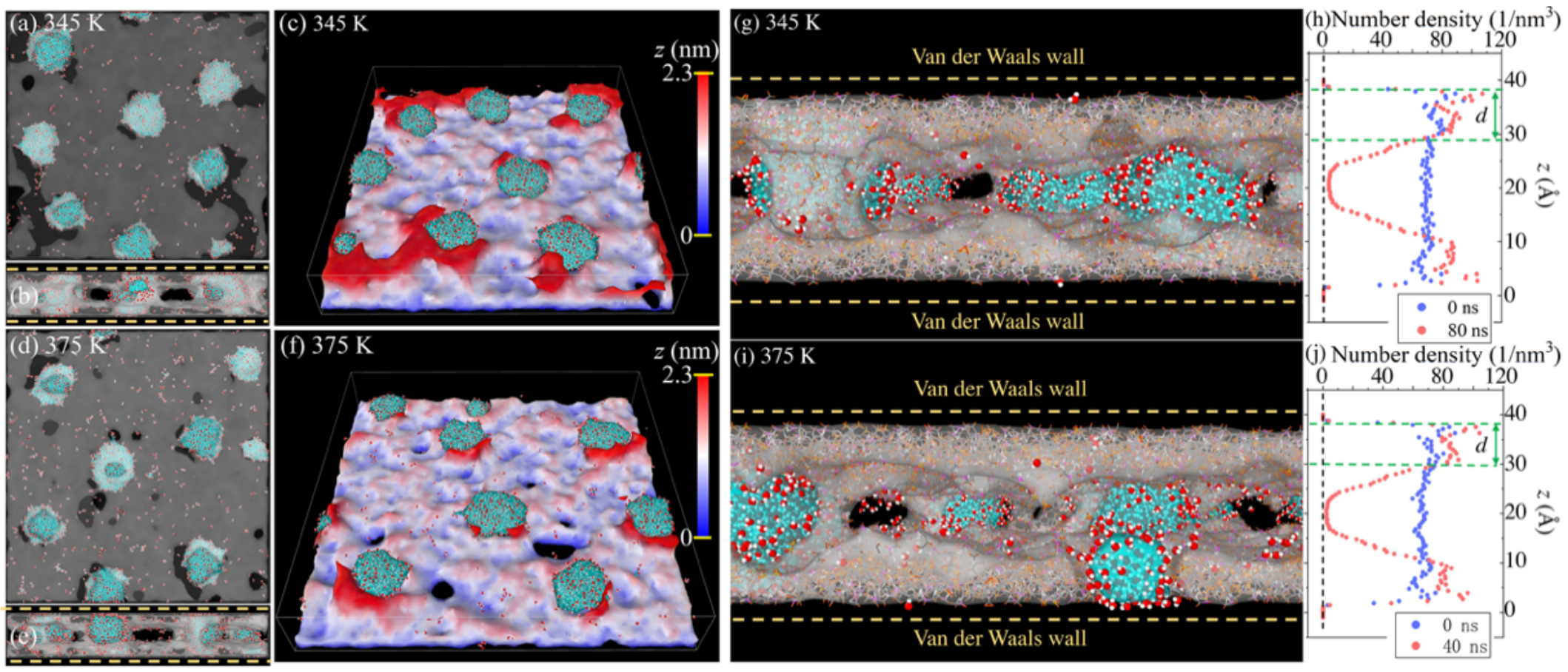

FIG. 4. The equilibrium pattern of the binary mixture at 345 K, top view (a), side view (b), and middle section (c), as well as corresponding patterns at 375 K (d-f). Organic nanochannels formed under vdW confinement and the distribution profile of resin at temperatures of 345 K (g, h) and 375 K (i, j).

To quantify the phase separation process, we obtained the evolution of distribution profiles of resin, as shown in Fig. 5(a, b). Then, we analyzed these profiles, deriving the number density of the middle region (*z* ranges from 1.8 to 2.2 nm) as a function of temporal power, as shown in Fig. 5(c). The analysis indicates a linear decrease of resin concentration ($C$) in the middle region with respect to temporal power ($t^{1/3}$), that is, $C(t,T) \propto a(T)t^{1/3}$, until the concentration approximately reaches 10%. In this case, $a(T)$ represents a temperature-dependent constant. Its values are $-18.58\,\mathrm{nm}^{-3}\cdot\mathrm{ns}^{-1/3}$ and $-28.33\,\mathrm{nm}^{-3}\cdot\mathrm{ns}^{-1/3}$ at temperatures of 345 K and 375 K, respectively. The separation velocity, as determined using a temporal power ($t^{1/3}$), exhibits a rapid reduction as the concentration diminishes beyond a specific threshold. This phenomenon is distinctly illustrated by the blue region in Fig. 5(c), with the critical point being approximately 14%. Beyond this critical point, we found the concentration exhibits a linear decrease with respect to time ($t$), expressed as $C(t,T) \propto t$, a relationship clearly demonstrated in Fig. 5(c). This observation aligns with the findings in literature [25], where surface tension is the dominating factor in the decay law at this stage.

To verify the universality of phase separation phenomena, we have expanded our simulation studies to include different water molecule models, system with homogeneous initial mixing, and system with a size expanded to four times the original. Their morphologies after 20 ns are shown in Supplementary Material Fig. S1. The separation laws for these models are shown in Fig. 5(d). After comparing the SPC, SPC-E, and the more parameter-rich TIP4P models, we observed that until the degree of separation dropped to 10%, the separation speed and power-law relationships of these models were essentially consistent with the TIP3P model. However, once the degree of phase separation in the homogeneous mixture of water and epoxy resin molecules exceeded 30%, their behavior significantly deviated from the previous power-law relationship (Details of the separation are shown in Fig. S2). It must be emphasized that achieving uniform molecular-level mixing of immiscible liquids is extremely difficult in practical applications, making our choice of initial state for the simulations, where a large block of water is in the middle of the resin, practically significant. Moreover, even when the size of the simulated system was expanded to four times larger, we did not observe any change in the separation laws. For more details on the simulations, see the supplementary materials. Taken together, our findings reveal that the phase separation phenomena and their power-law relationships reflect fundamental physical laws within nanoscale confined spaces.

Although the phase separation process can produce resin films, due to Rayleigh-like instability, the upper and lower layers of the resin film are prone to bridging and merging (see Fig. 6(a)). While vdW forces can mitigate this instability to a certain degree, additional mechanisms might be necessary to achieve the formation of stable nanochannels and liquid films. To elucidate the underlying mechanisms for addressing this challenge, we introduce a new concept termed "nano soft-architecture". As shown in Fig. 6(b), our simulations revealed how condensed water droplets can act as dynamic, movable "Pillars", thereby preventing the Rayleigh-like instability of the resin film. Since each water droplet "Pillar" can support only a limited area of resin film, this leads to the phenomenon shown in Fig. 1(h), where the resin cannot stably separate in most areas. These forces, $F_i$, can be calculated by integrating the resin/water interfacial tension, γ, along the circumferential contact line: $F_i = 2\pi r_i \gamma \mathrm{Sin}\theta_i$, as shown in Fig. 6(c). When the mixing ratio is appropriate, the droplets in the dynamic equilibrium system should be statistically uniformly distributed, meaning all droplets tend toward a consistent size, and the distance $d_{ij}$ between any two droplets tends toward stability, as shown in Fig. 4(a, d) and Fig. 6(b). These characteristics are related to the surface tension of the two liquids, the mixing ratio, the potential energy of the vdW wall, and the scale of the confined space, the specific impacts of which require further in-depth study.

To validate the hypothesis, we analyzed the model using the principle of virtual work based on *d'Alembert*'s principle, which enables us to treat dynamical problems as if they were statical [26]. Assuming that the water droplet moves vertically along the wall by a virtual displacement $\delta h$, the following relationship can be established:

$$\delta W = -\Pi S_i \delta h' + F_i \delta h \tag{1}$$

where $\Pi$ represents the disjoining pressure of the film and $S_i$ is the corresponding surface area. Assuming incompressibility, and taking into account symmetry and the neglect of higher-order infinitesimals, the approximate relationship $S_i \delta h' \approx \pi r_i^2 \delta h$ is obtained. Substituting this into Eq. (1) and setting $\delta W / \delta h = 0$, we obtain:

$$\Pi r_i = 2\gamma \mathrm{Sin}\theta_i \tag{2}$$

We approximated the pillar as spherical, leading to the relation:

$$\Pi R_i = 2\gamma \tag{3}$$

To minimize the influence of thermal fluctuations, we analyzed the data from the largest model, Scale-4, where the disjoining pressure in the z-direction is $279 \pm 20$ atm (see Fig. S3(a)). We determined the radius parameters $R_i \approx 1$ nm based on the resin's profile (see Fig. S3(b)). Consequently, the interfacial tension of DGEBA/water can be estimated form Eq (3) to be $14 \pm 1$ mN/m, which is close to the experimental value $17.5 \pm 0.1$ mN/m [27]. It is important to note that this experimental value was measured at room temperature, and the interfacial tension is expected to decrease slightly when the temperature is raised to 375 K [28]. Therefore, this analysis indirectly confirms the accuracy of the mechanism.

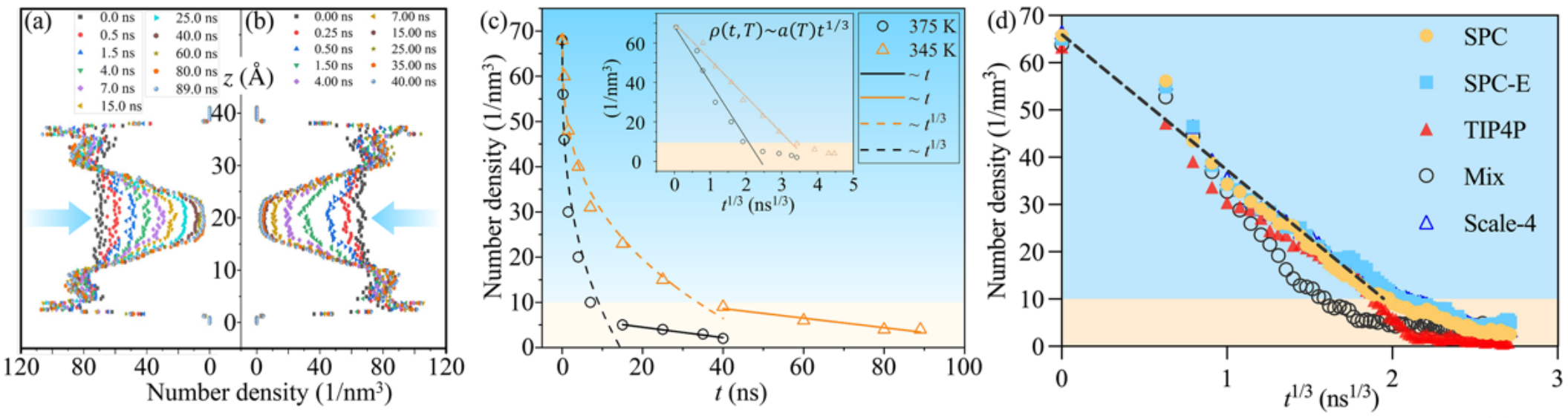

FIG. 5. The evolution of resin distribution profiles at temperatures of 345 K (a) and 375 K (b). (c) The density of the middle region as a function of temporal power (The insert clearly shows a linear relationship with the temporal power ($t^{1/3}$)). (d) Comparison of the separation process under different water molecule models, homogeneous mixing, and a model scaled up by four times with the TIP3P at 375 K (dashed line).

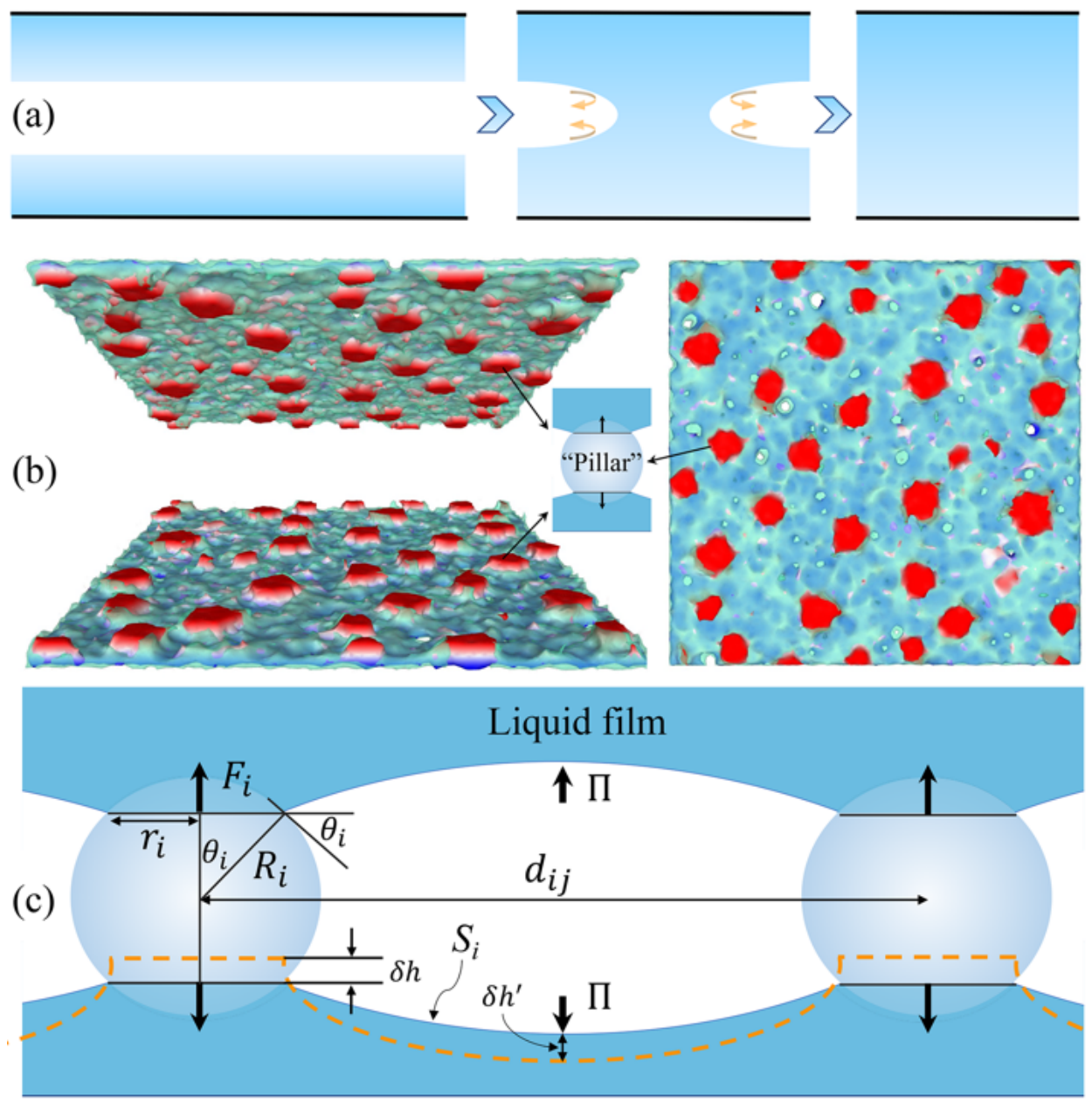

FIG. 6. Nano soft-architecture: (a) Challenges in forming stable liquid films due to Rayleigh-like instability. (b) At 375 K, the upper and lower film structures in the Scale-4 model at 20 ns and the "Pillar" (water droplets) supporting the films. (c) The mechanical mechanism that prevents the occurrence of Rayleigh-like instability.

*Summary*. This study presents a comprehensive analysis of the phase separation process in a water-epoxy resin system under vdW confinement, as examined through MD simulations. We identified a novel phenomenon, which we have termed PSTP, observable under specific temperature and concentration conditions. During PSTP, water molecules located centrally diffuse and then condense symmetrically. Concurrently, the epoxy resin separates along the thickness direction, leading to the formation of a nanochannel. Furthermore, our findings show that the rate of epoxy resin decay in the central region adheres to a temporal power law ($\propto t^{1/3}$) for approximately the first 86% of the decay process. Subsequently, this rate transitions to a slower linear relationship with time ($\propto t$). Finally, a new mechanism for the formation of stable nanochannels and films is proposed, namely the dynamic support of soft "Pillars" to prevent Rayleigh-like instability. These findings are fundamental to understanding the phase separations of binary mixtures under vdW confinement, and they offer potential applications in engineering fields.

Future research should be geared towards more comprehensive and systematic studies focusing on formation conditions. These include factors such as temperature, concentration, interactions with vdW walls, and the components of the binary liquid. Such conditions determine the physical quantities in the systems, including mobilities, disjoining pressures, surface tensions, viscosities, and diffusion coefficients. For instance, we can identify an appropriate polymer resin and curing agent to form liquid films through the PSTP process. Subsequent temperature increases will then trigger crosslinking reactions to form cured polymeric nanochannels. The width of these nanochannels and the thickness of the film can be manipulated by adjusting the water concentration, temperature, and parameters of the vdW confinement. The PSTP process, in conjunction with the concept of nano soft-architecture, enables the fabrication of biomimetic nanochannels [29], opening up potential future applications in fields ranging from energy conversion and separations to drug delivery. Thus, our findings may offer valuable insights and potential applications in the manufacture of nano-films and organic nanochannels, contributing significantly to the realms of bio-detection and energy.

I acknowledge the support from The Hong Kong Polytechnic University (Grant No. 1-W20C).